\documentclass[journal,final]{IEEEtran}

\usepackage{amsmath,graphicx,url}
\usepackage{amssymb}
\usepackage{math}
\usepackage{subfig}
\usepackage{algorithmic}
\usepackage{algorithm}
\usepackage{amsmath}

{}
{}
{}

\begin{document}

\title{Expectation-Maximization for Speech Source Separation Using Convolutive Transfer Function}

\author{Xiaofei Li, Laurent Girin and Radu Horaud
\thanks{X. Li and R. Horaud are with INRIA Grenoble Rh\^one-Alpes, Montbonnot Saint-Martin, France. }
\thanks{L. Girin is with GIPSA-lab and with Univ. Grenoble Alpes, Saint-Martin d'H\`eres, France.}
\thanks{This work was supported by the ERC Advanced Grant VHIA \#340113.}
}

\maketitle
\begin{abstract}
This paper addresses the problem of under-determinded speech source separation from multichannel microphone singals, i.e. the convolutive mixtures of multiple sources. The time-domain signals are first transformed to the short-time Fourier transform (STFT) domain. To represent the room filters in the STFT domain, instead of the widely-used narrowband assumption, we propose to use a more accurate model, i.e. the convolutive transfer function (CTF). At each frequency band, the CTF coefficients of the mixing filters and the STFT coefficients of the sources are jointly estimated by maximizing the likelihood of the microphone signals, which is resolved by an Expectation-Maximization (EM) algorithm. Experiments show that the proposed method provides very satisfactory performance under highly reverberant environments. 
\end{abstract}

\section{Introduction}\label{sec1}
Most of speech source separation techniques are designed in the short time Fourier transform (STFT) domain where the narrowband assumption is generally  used, e.g. [1-4]. In the narrowband assumption, at each frequency band, the time-domain filter is represented by the acoustic transfer function (ATF), and the time-domain convolutive process is transformed to a product between the ATF and the STFT coefficients of the source signal. This assumption is also referred to as the multiplicative transfer function (MTF) approximation [5]. Based on the ATF or its variant, e.g. relative transfer functions [6, 7],  beamforming techniques are widely used for multichannel speech source separation and speech enhancement. Popular beamformers for multisource separation include linearly constrained minimum variance/power (LCMV/LCMP) [6, 8]. Furtherly, because of the spectral sparsity of speech, the microphone signal can be assumed to be dominated by only one speech source in each time-frequency (TF) bin. This is refered to as the W-disjoint orthogonality (WDO) assumption [1].
The binary masking method [1, 2] and $\ell_1$-norm minimization method [3] exploit such WDO assumption.
More examples of narrowband assumption-based techniques can be found in [9] and references therein. 

In real scenarios, the time-domain filter, i.e. the room impulse response (RIR), is normally much longer than the STFT window (frame) length, since the latter should be set to be sufficiently short to account for the local stationarity of speech. For this case, the narrowband assumption is no longer valid, and thus leads to unsatisfied speech source separation performance. 
In the literature, only a few studies had questioned the validity of the narrowband assumption and  attempted to tackle this problem.    
Based on the narrowband assumption, the theoretical covariance matrix of one source image is a rank-one matrix [4]. To mitigate the invalidity of this matrix in practice,  a full-rank spatial covariance matrix was adopted in [10], even if the narrowband assumption is used. 
To circumvent the inaccurancy of the narrowband assumption, the wideband time-domain convolution model was used in [11, 12], where the source STFT coefficients are recovered by minimizing the fit cost between the time-domain mixtures and the time-domain convolution model. Meanwhile, based on the Lasso technique, the $\ell_1$-norm of the source STFT coefficients is minimized as well to impose the sparsity of speech spectra.  This method achieves good performance, but its computational complexity is very large due to the high cost of time-domain convolution operation. In [13, 14], based on the criterion of likelihood maximization, a variational EM algorithm was proposed also using the time-domain convolution model and the STFT domain signal model.

In [15], the time-domain convolution can be ideally represented in the STFT-domain by the cross-band filters. More precisely, the STFT coefficients of the source image can be computed by summing multiple convolutions (over frequencies) between the STFT coefficients of the source signal and the STTF-domain filter. Note that the convolution is conducted along the frame axis. To simplify the analysis, for each frequency, the band-to-band filter, i.e. the CTF model [16], is used, while with the cross-band information omitted. Compared to the narrowband assumption that uses a frequency-wise scalar product to approximate the time-domain convolution, the CTF model uses a frequency-wise convolution and thus is more accurate.
 Following the principle of the wide-band Lasso [11], based on the CTF model, a subband Lasso technique was proposed in [17], which largely reduces the complexity relative to the wide-band Lasso technique. In [18], two CTF inverse filtering methods were proposed based on the multiple-input/output inverse theorem (MINT). In [19], the CTF was integrated into the generalized sidelobe canceler beamformer. A CTF-based EM algorithm  was proposed in [20] for single-source dereverberation, in which the Kalman filter was exploited to achieve  online EM update. The cross-band filters were adopted in [21], combined with a non-negative matrix factorization model for the source signal. To estimate the source signals,  the likelihood of the microphone signals is maximized via a variational EM algorithm.  In [22], also based on likelihood maximization and EM, a STFT-domain convolutive model was used for source separation, combined with an HMM model for source activity. Even though this STFT-domain convolutive model was not named as CTF in [22], it actually plays the same role as CTF.  

Due to the high model complexity, the above mentioned source separation techniques that are beyond the narrowband assumption actually can not be performed in a blind manner, in other words, some prior knowledge are required. For example, the RIRs for the wide-band Lasso techniques, and the CTFs for the CTF-Lasso and MINT techniques, are required to be knwon or well estimated. Both the mixing filters and source parameters are required as a good initialization for the (variational) EM techniques [13, 14, 21, 22]. 
A blind multichannel CTF identification method was proposed in [23] and the identified CTF can be fed in the semi-blind methods. However, this CTF identification method was only suitable for the single-source case.

In the present paper, based on the CTF model, we propose a likelihood maximization method for speech source separation. First, the CTF model is presented in a source mixture probabilistic framework. Then, an EM algorithm is proposed for resovling the likelihood maximization problem. The STFT coefficients of the source signals are taken as hidden variables, and are estimated in the expectation step (E-step). The CTF coefficients and source parameters are estimated in the maximization step (M-step). 
Experiments show that the proposed method performs better than the narrowband assumpation based methods [1, 10] and the CTF-lasso method [17] within a semi-blind setup where the mixing filters are initialized with a perturbed version of the ground-truth CTF.   

The rest of this paper is organized as follows. Section~\ref{sec:ctf} presents the CTF formulation, which is plugged in a probabilistic framework in Section~\ref{sec:mass}. The proposed EM is given in Section~\ref{sec:em}. Experiments are presented in Section~\ref{sec:exp}. Section~\ref{sec:con} concludes the paper. 
This paper is an extension of a conference paper [24]. The main improvements over [24] consists of i) we present the methodology in more detail, such as the two vector/matrix formulations in Section \ref{sec:mass},   the detailed derivation of the EM algorithm in Section \ref{sec:em} and the execution process of EM in Algorithm 1; ii) in Section \ref{sec:exp}, we add the experiments with CTF perturbations, and analyze the computational complexity of the proposed method.

\section{CTF Formulation}
\label{sec:ctf}

In an indoor (room) environment, a speech source signal propagates to the receivers (microphones) through the room effect. In the time domain, the received source image ${y}(n)$ is given by the linear convolution between 
the speech source signal ${s}(n)$ and the RIR ${a}(n)$:
\begin{align}\label{eq:yn}
 y(n) =  a(n) \star s(n), 
 \end{align} 
 where $\star$ denotes convolution. Applying STFT, the narrowband approximation  usually approximates the time domain convolution (\ref{eq:yn}) as the product $y(p,k)=a(k)  s(p,k)$, where $y(p,k)$ and $s(p,k)$ are the STFT coefficients of the corresponding signals, and $a(k)$ is the ATF. Let $N$ denote the STFT window (frame) length, then $k \in[0,N-1]$ is the frequency bin index, and $p \in[1,P]$ is the frame index.
This narrowband approximation is no longer valid when ${a}(n)$ is long relative to the STFT window. 

We use the CTF model to circumvent the inaccurancy of the narrowband assumption, then $y(p,k)$ can be presented as [16]:
\begin{align}\label{eq:xpk3}
 y(p,k) =  a(p,k) \star s(p,k) = \sum_{p'=0}^{Q} a(p',k) s(p-p',k),  
 \end{align}
where $a(p,k)$ denotes the CTF, which can be derived from the time-domain filter ${a}(n)$ by:
\begin{align}\label{eq:apk}
a(p,k)=({a}(n) \star \zeta_{k}(n))|_{n=pL}.  
\end{align}
This equation is the convolution with respect to the time index $n$ evaluated at multiples of the frame step $L$, with
\begin{align}
\zeta_{k}(n) = e^{j\frac{2\pi}{N}kn}\sum_{m=-\infty}^{+\infty} \tilde{\omega}_a(m) \: \tilde{\omega}_s(n+m), \nonumber
\end{align}
where $\tilde{\omega}_a(n)$ and $\tilde{\omega}_s(n)$ are respectively the STFT analysis and synthesis windows.
The length of CTF, i.e. $Q+1$, approximately equals the length of RIR divided by $L$.

\section{Mixture Model Formulations}
\label{sec:mass}

\subsection{Basic Formuation for Mixture Model}\label{sec:pm}

We consider a source separation problem with $J$ sources and $I$ sensors, which could be either underdetermined ($I<J$) or (over)determined ($I\ge J$). Using the CTF formulation (\ref{eq:xpk3}), in the STFT domain, the microphone signal $x_i(p,k)$ is:
\begin{align}\label{eq:basicmix}
 x_i(p,k) = \sum_{j=1}^J a_{ij}(p,k) \star s_j(p,k) + e_i(p,k),  
\end{align}
where $a_{ij}(p,k)$ is the CTF from source $j, \ j=1,\dots,J$ to sensor $i, \ i=1,\dots,I$, and $e_i(p,k)$ denotes the noise signal.

\subsection{Probabilistic Model}
In the literature of source separation, each source signal $s_j(p,k)$ is normally assumed to be independent to  other sources, and is also  independent across STFT frames and frequencies. Each STFT coefficient $s_j(p,k)$ is assumed to follow a complex Gaussian distribution with a zero mean and variance $v_j(p,k)$ [4, 10], 
i.e. its probability density function (pdf) is:
\begin{align}
\mathcal{N}_c(s_j(p,k);0,v_j(p,k))=\frac{1}{\pi v_j(p,k)}\exp \big(\!-\!\frac{|s_j(p,k)|^2}{v_j(p,k)} \big). \nonumber
\end{align}

The noise signal is assumed to be stationary, temporally uncorrelated, and  independent to the speech source signals. We define the noise vector accross microphones as $\evect(p,k) =  [e_1(p,k),\dots,e_I(p,k)]^{\top} \in \mathbb{C}^{I\times 1}$. This vector is assumed to follow a zero-mean complex Gaussian, with a non-diagonal covariance matrix denoted as $\mathbf{\Sigma_e}(k)$. This convariance matrix encodes the spatial correlation of the noise signals.  The pdf is:
\begin{align}
\mathcal{N}_c(\evect(p,k);0,\mathbf{\Sigma_e}(k)) = \frac{1}{\pi^I |\mathbf{\Sigma_e}(k)|} e^{-\evect(p,k)^H\mathbf{\Sigma_e}(k)^{-1}\evect(p,k)}, \nonumber
\end{align}
where $^H$ denotes complex transpose, $|\cdot|$ the determinant of matrix.

Since the proposed source separation method is carried out independently at each frequency, hereafter, we omit the frequency index $k$ for notational simplicity.

\subsection{Vector/matrix Formulation 1}
To formulate the mixture model (\ref{eq:basicmix}) more compactly,  we have several different choices to organize the signals and the convolution operation in vector/matrix forms. To facilitate the derivation of the following EM algorithm, we use two different vector/matrix formulations. In this section, Formulation 1 will be presented to enable us to easily derive the M-step,  and Formulatioin 2 will be presented used for the derivation of the E-step in the next section. This two formulations are different just in the organization of the variables and parameters, thence transforming from the E-step to the M-step, and vice-versa, will only necessitate reorganizing the vector/matrix elements.

In Formulation 1, we define the source signals in vector form as, for $p \in [1,P]$: 
\begin{align}
 \svect_j(p)&=[s_j(p),\dots,s_j(p-q),\dots,s_j(p-Q)]^{\top} \in \mathbb{C}^{(Q+1)\times 1}, \nonumber \\
 \svect(p)&=[\svect_1(p)^{\top},\dots,\svect_j(p)^{\top},\dots,\svect_J(p)^{\top}]^{\top} \in \mathbb{C}^{J(Q+1)\times 1}, \nonumber
\end{align}
where $^{\top}$ denotes vector/matrix transpose. If $p \leq q$, we set $s_j(p-q)=0$. Define the CTF in vector/matrix form as: 
\begin{align}
\mathbf{a}_{ij}&=[a_{ij}(0),\dots,a_{ij}(p),\dots,a_{ij}(Q)]^{\top}\in\mathbb{C}^{ (Q+1)\times 1}, \nonumber \\
\mathbf{A}_j&=[\mathbf{a}_{1j}^{\top};\dots;\mathbf{a}_{ij}^{\top};\dots;\mathbf{a}_{Ij}^{\top}]\in\mathbb{C}^{I\times (Q+1)}, \nonumber \\
\mathbf{A}&=[\mathbf{A}_1,\dots,\mathbf{A}_j,\dots,\mathbf{A}_J]\in\mathbb{C}^{I\times J(Q+1)}, \nonumber
\end{align}
We already defined $\evect(p) \in \mathbb{C}^{I\times 1}$. Similarly, the microphone signal is $\xvect(p,k) =  [x_1(p,k),\dots,x_I(p,k)]^{\top} \in \mathbb{C}^{I\times 1}$. Finally, we can rewrite (\ref{eq:basicmix}) as:
 \begin{align}\label{eq:x}
 \xvect(p) = \sum_{j=1}^J \mathbf{A}_j \svect_j(p)+\evect(p)=\mathbf{A} \svect(p)+\evect(p).  
 \end{align}
 
In Formulation 1, the source vector $\svect(p)$ follows a zero-mean complex Gaussian distribution with $J(Q+1) \times J(Q+1)$ diagonal covariance matrix $\mathbf{R_s}(p)$ where the first $Q+1$ diagonal entries (for the first source) are $v_1(p),\dots,v_1(P-Q)$, the next $Q+1$ diagonal entries (for the second source) are $v_2(1),\dots,v_2(P)$, and so on.
The pdf of the mixture given the sources is $\mathcal{N}_c(\xvect(p);\mathbf{A} \svect(p),\mathbf{\Sigma_e})$.

\subsection{Vector/matrix Formulation 2}

In {Formulation 2}, let 
\begin{align}
\tilde{\svect}_j &= [s_j(1),\dots,s_j(p),\dots,s_j(P)]^T\in\mathbb{C}^{ P\times 1}, \nonumber \\
\tilde{\evect}_i &= [e_i(1),\dots,e_i(p),\dots,e_j(P)]^T\in\mathbb{C}^{ P\times 1}, \nonumber \\
\tilde{\xvect}_i &= [x_i(1),\dots,x_i(p),\dots,x_j(P)]^T\in\mathbb{C}^{ P\times 1} \nonumber
\end{align}
 denote the $j$-th source vector and the $i$-th noise and microphone vectors, all involving all $P$ frames. Concatenate them along source/microphone as:
 \begin{align}
  \tilde{\svect} &= [\tilde{\svect}_1^{\top},\dots,\tilde{\svect}_j^{\top},\dots,\tilde{\svect}_J^{\top}]^{\top}\in\mathbb{C}^{JP\times 1}, \nonumber \\
   \tilde{\evect} &= [\tilde{\evect}_1^{\top},\dots,\tilde{\evect}_i^{\top},\dots,\tilde{\evect}_I^{\top}]^{\top}\in\mathbb{C}^{IP\times 1}, \nonumber \\
 \tilde{\xvect} &= [\tilde{\xvect}_1^{\top},\dots,\tilde{\xvect}_i^{\top},\dots,\tilde{\xvect}_I^{\top}]^{\top}\in\mathbb{C}^{IP\times 1}. \nonumber
  \end{align}
Define the CTF convolution matrix in $\mathbb{C}^{P\times P}$: 
\begin{equation}
\mathbf{\mathcal{A}}_{ij} =
\begin{bmatrix}
a_{ij}(0) & 0     &  \cdots & \cdots& \cdots   & 0  \\
\vdots & \ddots & \ddots& \ddots & \ddots  & \vdots \\
a_{ij}(Q)  & \ddots & a_{ij}(0) & 0 & \ddots  & 0 \\
\vdots & \ddots & \ddots & \ddots & \ddots   & \vdots \\
0& \cdots &0 & a_{ij}(Q) & \cdots & a_{ij}(0) \\
\end{bmatrix},  \nonumber 
\end{equation}
where the  CTF $\{a_{ij}(p)\}$ is first flipped and then duplicated as the row vectors, with one element shift per row. Concatenate it along source and microphone as:
\begin{align}\label{eq:conmat}
\mathbf{\mathcal{A}}_{i} &= [\mathbf{\mathcal{A}}_{i1}, \dots,\mathbf{\mathcal{A}}_{ij}, \dots, \mathbf{\mathcal{A}}_{iJ}]\in\mathbb{C}^{P\times JP}, \nonumber \\
\mathbf{\mathcal{A}} &= [\mathbf{\mathcal{A}}_{1}^{\top},\dots,\mathbf{\mathcal{A}}_{i}^{\top},\dots,\mathbf{\mathcal{A}}_{I}^{\top}]^{\top}\in\mathbb{C}^{IP\times JP}. 
\end{align}
Then we can rewrite (\ref{eq:basicmix}) as: 
\begin{align}
 \tilde{\xvect}_i &= \sum_{j=1}^J\mathbf{\mathcal{A}}_{ij}\tilde{\svect}_j +\tilde{\evect}_i= \mathbf{\mathcal{A}}_{i}\tilde{\svect}+\tilde{\evect}_i,  \nonumber \\
  \text{or} \qquad \tilde{\xvect}&=\mathbf{\mathcal{A}}\tilde{\svect}+\tilde{\evect}. 
\end{align}

In Formulation 2, the pdf of source vector $\tilde{\svect}$ is a zero-mean complex Gaussian distribution with $JP \times JP$ diagonal covariance matrix 
\begin{align}\label{eq:diag}
\mathbf{\Psi_s} = \text{Diag}\{[v_1(1),\dots,v_1(P),\dots,v_J(1),\dots,v_J(P)]^{\top}\}
\end{align} 
 where $\text{Diag}\{\cdot\}$ denotes diagonal matrix of a vector. The noise vector $\tilde{\evect}$ follows a zero-mean complex Gaussian distribution with $IP \times IP$ covariance matrix $\mathbf{\Psi_e}$. The entries of  $\mathbf{\Psi_e}$ are
\begin{equation}\label{eq:Psie}
\mathbf{\Psi_e}((i_1-1)P+p_1,(i_2-1)P+p_2) = 
\begin{cases} \mathbf{\Sigma_e}(i_1,i_2), & \mbox{if } p_1= p_2, \\
0, &\mbox{otherwise.}
 \end{cases}
\end{equation}
where the arguments in parentheses denotes the row and column indices; $i_1,i_2 \in [1,I], \ p_1,p_2 \in [1,P]$. The pdf of the mixture given the sources is $\mathcal{N}_c(\tilde{\xvect};\mathbf{\mathcal{A}} \tilde{\svect},\mathbf{\Psi_e})$.

\section{Expectation-Maximization Algorithm}
\label{sec:em}

Collect the source variances for all sources and frames, we have the source variance set $\mathbf{V}=\{v_j(p)\}_{j \in [1,J],p \in [1,P]}$. The parameters set in the present problem is $\Theta=\{\mathbf{V},\Amat,\mathbf{\Sigma_e}\}$ in Formulation 1 or $\Theta=\{\mathbf{\Psi_s},\mathbf{\mathcal{A}},\mathbf{\Psi_e}\}$ in Formulation 2. 
The likelihood of the mixture will be maximized by an EM 
algorithm, in which the parameters $\Theta$ are the optimization variables. Meanwhile, the the STFT coefficients of the source signals, i.e. $\{s_j(p)\}_{j,p}$, are taken as hidden variables, whose posterior statistics will be inferred, and the posterior mean is taken as the estimation of the source signals. The proposed EM algorithm is summarized in Algorithm~\ref{alg:em}.

\subsection{E-step}

The E-step will be derived based on Formulation 2. Using the parameters estimates $\Theta$, given in the preceding M-step by (\ref{eq:Anew}) in Formulation 1, we construct the CTF convolution matrix $\mathbf{\mathcal{A}}$, the source covariance matrix $\mathbf{\Psi_s}$   and the noise covariance matrix $\mathbf{\Psi_e}$ following (\ref{eq:conmat}), (\ref{eq:diag}) and (\ref{eq:Psie}), respectively.

In Formulation 2, the posterior distribution of the source signals is $p(\tilde{\svect}|\tilde{\xvect},\Theta) \propto p(\tilde{\xvect}|\tilde{\svect},\Theta)p(\tilde{\svect}|\Theta)$. Since both $p(\tilde{\xvect}|\tilde{\svect},\Theta)$ and $p(\tilde{\svect}|\Theta)$ are Gaussian,  $p(\tilde{\svect}|\tilde{\xvect},\Theta)$ is also Gaussian. Let $\mathbb{E}_{\tilde{\svect}|\tilde{\xvect},\Theta}[ \ . \ ]$ denotes the expectation in the sense of the posterior distribution $p(\tilde{\svect}|\tilde{\xvect},\Theta)$. From the exponent of $p(\tilde{\svect}|\tilde{\xvect},\Theta)$, i.e. 
\begin{align}
-(\tilde{\xvect}-\mathcal{A} \tilde{\mathbf{s}})^H \mathbf{\Psi_e}^{-1}(\tilde{\xvect}-\mathcal{A} \tilde{\mathbf{s}}) - \tilde{\svect}^H \mathbf{\Psi_s}^{-1} \tilde{\svect},  \nonumber 
\end{align} 
the posterior mean $\widehat{\tilde{\svect}}= \mathbb{E}_{\tilde{\svect}|\tilde{\xvect},\Theta}[\tilde{\svect}]$ and covariance matrix $\mathbf{\widehat{\Sigma}_s}=\mathbb{E}_{\tilde{\svect}|\tilde{\xvect},\Theta}[(\tilde{\svect}-\widehat{\tilde{\svect}})(\tilde{\svect}-\widehat{\tilde{\svect}})^{H}]$ can be derived by reorganizing the quadratic and linear forms in $\tilde{\svect}$ .
We obtain: 
\begin{align}\label{eq:poss}
\mathbf{\widehat{\Sigma}_s} &= \big( \mathbf{\mathcal{A}}^H \mathbf{\Psi_e}^{-1} \mathbf{\mathcal{A}}+ \mathbf{\Psi_s}^{-1} \big)^{-1}, \nonumber \\
\widehat{\tilde{\svect}} &= \mathbf{\widehat{\Sigma}_s} \mathbf{\mathcal{A}}^H \mathbf{\Psi_e}^{-1} \tilde{\xvect},  
\end{align}
and then the posterior second-order moment matrix $\mathbf{\widehat{\tilde{R}}_s}= \mathbb{E}_{\tilde{\svect}|\tilde{\xvect},\Theta}[\tilde{\svect}\tilde{\svect}^H]$ can be computed as 
\begin{align}\label{eq:som}
\mathbf{\widehat{\tilde{R}}_s}=\widehat{\tilde{\svect}}\widehat{\tilde{\svect}}^{H}+\mathbf{\widehat{\Sigma}_s}.
\end{align}
As derived based on the narrowband assumption in [4],  Eq.~(\ref{eq:poss}) is also in the form of classical Wiener filtering. 
The difference is, here the interframe elements in the posterior covariance matrix $\mathbf{\widehat{\Sigma}_s}$ are nonzero, which means the correlation between frames due to the convolution is encoded. As a result, the posterior mean of source, i.e. $\widehat{\tilde{\svect}}$, is recovered by deconvoluting the mixture.

\subsection{M-step}

The M-step will be derived based on Formulation 2. Collect the multichannel mixture vectors and source signal vectors along frames, we have the observation set  $\Xmat=\{\xvect(p)\}_{p \in [1,P]}$ and the source signal set $\Smat=\{\svect(p)\}_{p \in [1,P]}$. The complete-data (including observations and hidden variables) likelihood function is:
\begin{align}
p(\Xmat,&\Smat|\Theta) \propto p(\Xmat|\Smat,\Theta)p(\Smat|\Theta) \nonumber \\
&=\prod_{p=1}^P\mathcal{N}_c(\xvect(p);\mathbf{A} \svect(p),\mathbf{\Sigma_e})\prod_{j=1}^J\prod_{p=1}^P\mathcal{N}_c(s_j(p);0,v_j(p)). \nonumber
\end{align}
 Considering only the terms related to the parameters and hidden variables, the corresponding loglikelihood writes:
 \begin{align}\label{eq:loglik}
 &\log(p(\Xmat,\Smat|\Theta)) = - \sum_{p=1}^P \Big(\text{log}(|\mathbf{\Sigma_e}|)+(\xvect(p)-\mathbf{A} \svect(p))^H\mathbf{\Sigma_e}^{-1}\nonumber \\ & (\xvect(p)-\mathbf{A} \svect(p))\Big)  - \sum_{j=1}^J\sum_{p=1}^P \Big( \log(v_j(p)) + \frac{|s_j(p)|^2}{v_j(p)} \Big) + \text{const}. 
 \end{align}
Denote the auxiliary function for likelihood maximization as  $Q(\Theta,\Theta^{\text{old}}) = \mathbb{E}_{\mathbf{S}|\mathbf{X},\Theta^{\text{old}}}[\log(p(\Xmat,\Smat|\Theta))]$, where $\Theta^{\text{old}}$ denotes the parameters estimated at the previous iteration. From the loglikelihood (\ref{eq:loglik}), the auxiliary function can be derived as:
\begin{align}
Q(\Theta,\Theta^{\text{old}}) = &-\sum_{p=1}^P\Big(\text{log}(|\mathbf{\Sigma_e}|)+ 
\text{Trace}\{\mathbf{\Sigma_e}^{-1}(\mathbf{A}\widehat{\svect}(p)\xvect(p)^H \nonumber \\ & \qquad \qquad+ \xvect(p)\widehat{\svect}(p)^H\mathbf{A}^H  
 - \mathbf{A}{\mathbf{\widehat{R}_s}(p)}\mathbf{A}^H)\} \Big) \nonumber \\
&- \sum_{j=1}^J\sum_{p=1}^P \Big( \log(v_j(p)) + \frac{\widehat{v}_j(p)}{v_j(p)} \Big) + \text{const},  
\end{align}
where $\text{Trace}\{\cdot\}$ denotes matrix trace, and 
\begin{align}
\widehat{\svect}(p) &= \mathbb{E}_{\mathbf{S}|\mathbf{X},\Theta^{\text{old}}}[ \svect(p)], \nonumber \\
\mathbf{\widehat{R}_s}(p) &= \mathbb{E}_{\mathbf{S}|\mathbf{X},\Theta^{\text{old}}}[\svect(p)\svect(p)^H], \nonumber \\
\widehat{v}_j(p) &= \mathbb{E}_{\mathbf{S}|\mathbf{X},\Theta^{\text{old}}}[|s_j(p)|^2] \nonumber 
\end{align}
are the posterior statistics of the source signal, namely the posterior mean, the posterior second-order moment matrix and the element-wise posterior second-order moment, respectively.  Actually, $\widehat{v}_j(p)$  is the $((j-1)(Q+1)+1)$-th diagonal entry of $\mathbf{\widehat{R}_s}(p)$.

\textbf{Reformulation}: $\widehat{\svect}(p)$ and $\mathbf{\widehat{R}_s}(p)$ can be obtained by reformulating $\widehat{\tilde{\svect}}$ and $\mathbf{\widehat{\tilde{R}}_s}$ derived in the preceding E-step. The reformulation is mainly to find the elements with the same source and frame indices:
\begin{align}\label{eq:refor}
\widehat{\svect}(p)_{\{j,p-q\}} &= \widehat{\tilde{\svect}}_{\{j,p-q\}} \nonumber \\
 \mathbf{\widehat{R}_s}(p)_{\{j_1,p-q_1\},\{j_2,p-q_2\}} &= \mathbf{\widehat{\tilde{R}}_s} \ _{\{j_1,p-q_1\},\{j_2,p-q_2\}},
\end{align}
where the subscript $_{\{j,p\}}$ denote ``the $j$-th source at $p$-th frame''.  In a vector, or in a row/column of a matrix, $_{\{j,p\}}$ represents i) the $\{(j-1)P+p\}$-th element in $\widehat{\tilde{\svect}}$ and $\mathbf{\widehat{\tilde{R}}_s}$, based on Formulation 2, and ii) the $\{(j-1)(Q+1)+q+1\}$-th ($q\in[0,Q]$) element in $\widehat{\svect}(p)$  and $\mathbf{\widehat{R}_s}(p)$, based on Formulation 1.

With respect to $\mathbf{A}^*$ ($^*$ denotes conjugate), $v_j(p)$ and $\mathbf{\Sigma_e}$, the (complex) derivative of $Q(\Theta,\Theta^{\text{old}})$  are respectively:
 \begin{align}\label{eq:dqa}
 \frac{\partial Q(\Theta,\Theta^{\text{old}})}{\partial \mathbf{A}^*} &=- \mathbf{\Sigma_e}^{-1}\sum_{p=1}^P \Big( \xvect(p)\widehat{\svect}(p)^H-\mathbf{A}{\mathbf{\widehat{R}_s}(p)} \Big), \nonumber \\
 \frac{\partial Q(\Theta,\Theta^{\text{old}})}{\partial v_j(p)} &=  \widehat{v}_j(p)v_j^{-2}(p) - v_j^{-1}(p), \nonumber \\  
 \frac{\partial Q(\Theta,\Theta^{\text{old}})}{\partial \mathbf{\Sigma_e}} &= \sum_{p=1}^P\Big(\mathbf{\Sigma_e}^{-1}(\mathbf{A}\widehat{\svect}(p)\xvect(p)^H + \xvect(p)\widehat{\svect}(p)^H\mathbf{A}^H  
\nonumber \\ & \qquad \qquad - \mathbf{A}{\mathbf{\widehat{R}_s}(p)}\mathbf{A}^H)\mathbf{\Sigma_e}^{-1}-\mathbf{\Sigma_e}^{-1}\Big)
 \end{align}
To maximize $Q(\Theta,\Theta^{\text{old}})$, the three derivatives are set equal to zero, then  $\mathbf{A}$, $v_j(p)$ and $\mathbf{\Sigma_e}$ can be estimated as, respectively:
\begin{align}\label{eq:Anew}
\mathbf{A}^{\text{new}} &= \Big(\sum_{p=1}^P \xvect(p)\widehat{\svect}(p)^H \Big) \Big(\sum_{p=1}^P \mathbf{\widehat{R}_s}(p) \Big)^{-1},  \nonumber \\
v_j^{\text{new}}(p) &= \widehat{v}_j(p). \nonumber \\
\mathbf{\Sigma_e}^{\text{new}} & = \frac{1}{P}\sum_{p=1}^P\Big(\mathbf{A}^{\text{new}}\widehat{\svect}(p)\xvect(p)^H + \xvect(p)\widehat{\svect}(p)^H\mathbf{A}^{\text{new} \ H}  \nonumber \\ & \qquad \qquad \qquad \qquad
 - \mathbf{A}^{\text{new}}{\mathbf{\widehat{R}_s}(p)}\mathbf{A}^{\text{new} \ H}\Big)
\end{align}

\begin{algorithm}[t]  \caption{\label{alg:em} EM for MASS with CTF} 
\begin{algorithmic} 
 \STATE Input: $\{x_i(p,k)\}_{p \in [1,P], k \in [0,N-1]}$; initial parameters $\Theta$.
 \REPEAT
 \STATE \textbf{E-step}
 \STATE 1 Construct $\mathbf{\mathcal{A}}$, $\mathbf{\Psi_s}$ and $\mathbf{\Psi_e}$ following (\ref{eq:conmat}), (\ref{eq:diag}) and (\ref{eq:Psie}), respectively,  
 \STATE 2 Compute $\mathbf{\widehat{\Sigma}_s}$ and $\widehat{\tilde{\svect}}$ following~(\ref{eq:poss}),
 \STATE 3 Compute $\mathbf{\widehat{\tilde{R}}_s}$ following~(\ref{eq:som}),
 \STATE \textbf{M-step}
 \STATE 4 Construct $\widehat{\svect}(p)$ and $\mathbf{\widehat{{R}}_s}(p)$ following~(\ref{eq:refor}),
 \STATE 5 Update $\mathbf{A}$, $v_j$ and $\mathbf{\Sigma_e}$  following~(\ref{eq:Anew}), 
 \UNTIL convergence
 \STATE Output: STFT coefficients of source signals $\widehat{\tilde{\svect}}$.
\end{algorithmic}
\end{algorithm} 

\section{Experiments}
\label{sec:exp}

\subsection{Experimental Configuration}

\begin{figure*}[t]
\centering
{\includegraphics[width=0.31\textwidth]{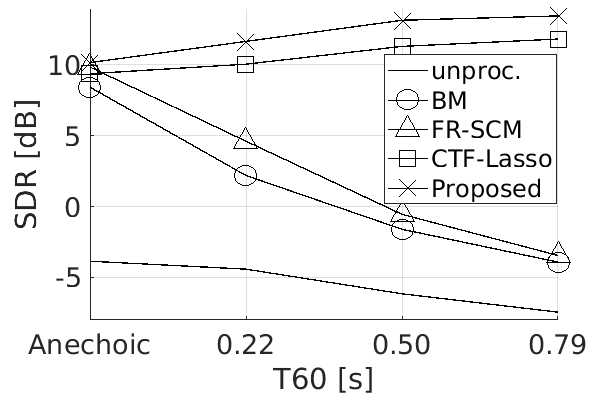}}
{\includegraphics[width=0.31\textwidth]{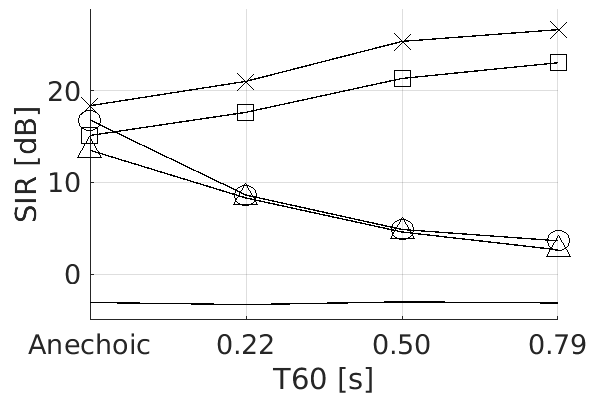}} 
{\includegraphics[width=0.31\textwidth]{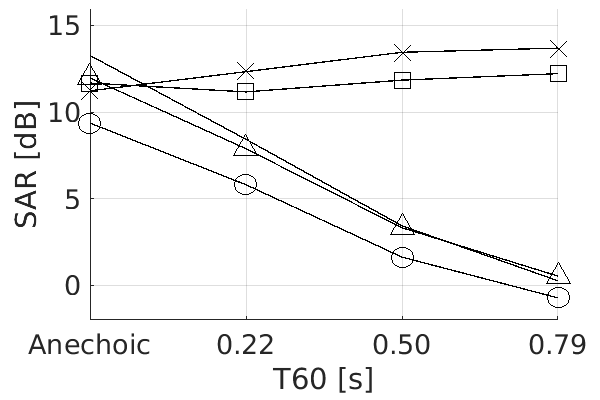}} 
\caption{Performance measures  as a function of $T_{60}$. The number of sources is three.  NPM is -35 dB. 'unproc.' represents the unprocessed mixture signals. } 
\label{fig:t60}
\vspace{-0.0cm}
\end{figure*}

The binaural (two-channel) simulated signals were used to evaluate the proposed EM algorithm. The experiments were conducted under various acoustic conditions, in terms of reverberation time, number of sources and intensity of room filter perturbation.

\subsubsection{Simulation set-up} We use a KEMAR dummy head [25] with one microphone embedded in each ear as the recording system. The head related impulse responses (HRIRs) for a large grid of directions were measured in advance. The ROOMSIM simulator [26] simulates the binaural room impulse responses using these HRIRs as both the direct-path wave and the reflections. Four reverberant conditions were simulated, with the reverberation time $T_{60}$ as $0$~s (the anechoic case), $0.22$~s, $0.5$~s and $0.79$~s, respectively.
The TIMIT [27] speech signals were used as the speech source signals, and were convolved with the simulated BRIRs to generate microphone signals (mixtures). The sampling rates of source signals and microphone signals are both 16-kHz, and the length of source signals are about 3 s. 
The speech sources were set to locate at different directions in front of the dummy head. 
The noisy microphone signals is generated by adding a spatially uncorrelated stationary speech-like noise to the noise-free signals. One SNR (signal-to-noise ratio) condition,  i.e. 20 dB, is tested. 
The STFT uses Hamming window with a length of $1,\!024$ samples ($64$~ms) and frame step of $256$ samples ($16$-ms).
In this experiment, the ground-truth noise covariance matrix $\mathbf{\Sigma_e}(k)$ is used, and fixed during EM iterations.

\subsubsection{EM initialization} 
Depends on what types of prior knowledges are available, the EM algorithm can be initialized either from the E-step or from the M-step. For both choices, the accuracy of the initialization is crucial for the EM iterations to converge to a good solution. In this experiment, we initialize EM from the M-step.
Due to the difficulty  of the blind initialization,  we consider a semi-blind initialization scheme. The time-domain filters are assumed to be known, from which the CTFs are computed by (\ref{eq:apk}). 
To fit the realistic situation that the time-domain filters (or CTFs)  should actually be blindly estimated and suffer from some estimation error,   
a proportional Gaussian random noise is added to the time-domain filters to generate the perturbed filters.
The normalized projection misalignment (NPM) [28] in decibels (dB) is used to measure the intensity of the perturbation. The lower NPM is, the intenser the perturbation will be. 
To have a good initialization for the source variance, the CTF-Lasso method proposed in [17] is first applied, which solves the  problem: 
\begin{align}\label{eq:cvx}
\mathop{\textrm{min}}_{\svect} \parallel \mathbf{\mathcal{A}} \tilde{\svect} - \tilde{\xvect} \parallel_2^2+\lambda \| \tilde{\svect} \|_1, \nonumber
\end{align}
where $\| \cdot\|_1$ denotes $\ell_1$-norm, and is used to impose the spectral spasity of speech sources. For more details, please refer to [17]. Then the source variance is initialized as the magnitude square of each source coefficient estimate. It is found that, for most of acoustic conditions and frequency bins, the number of EM iterations required for convergency is less than 10, thence in this experiment we use a constant number, i.e. 7, of EM iterations. 
 
\subsubsection{Baseline methods} Three baseline methods were used for comparison:  i) The CTF-Lasso method used for initialization; ii) The binary masking (BM) method [1], which is based on the narrowband approximation. 
To make a fair comparison, the narrowband mixing filters are also computed using the knwon perturbed time-domain filters. However, to compute the one-order mixing filters, for the high reverberation time cases, the time-domain filters should be first truncated to have a length being equal to (or less than) the STFT window length. Based on some pilot experiments, we use the HRIRs (without reverberation) as the truncated filters, which achieves the best results. For source separation, each TF bin is assigned to one of the sources   based on the mixing filters;  iii) the full-rank spatial covariance matrix (FR-SCM) method [10]. The full-rank spatial covariance matrix for each source was separately estimated using the corresponding source image, and kept fixed during the EM, following the line of the semi-oracle experiments in [10].

\subsubsection{Performance metrics} The signal-to-distortion ratio (SDR), signal-to-interferences ratio (SIR) and signal-to-artifacts ratio (SAR) [29], all in decibels (dB), are used as the separation performance metric. 
In the following, three sets of experiments will be conducted: i) for various reverberation time, ii) for various numbers of sources, i.e. with $2$, $3$, $4$ and $5$ sources, iii) with various NPM settings. For each condition, the  metric scores are averaged over 20 mixtures. 

The computation complexity of each method is measured with the real-time factor, which is the processing time of one method divided by the length of the processed signal. Note that all the methods are implemented in MATLAB.

\subsection{Results as a Function of Reverberation Time}

\begin{figure*}[t]
\centering
{\includegraphics[width=0.31\textwidth]{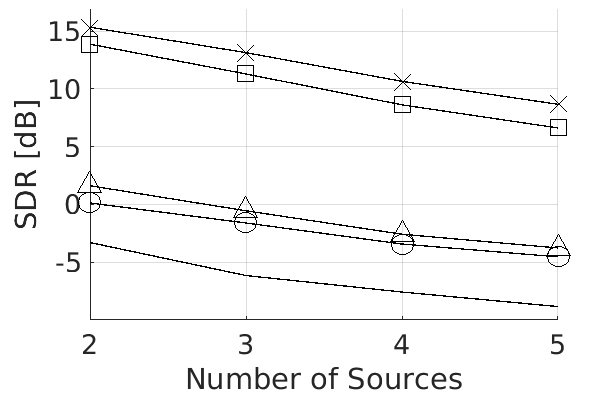}}
{\includegraphics[width=0.31\textwidth]{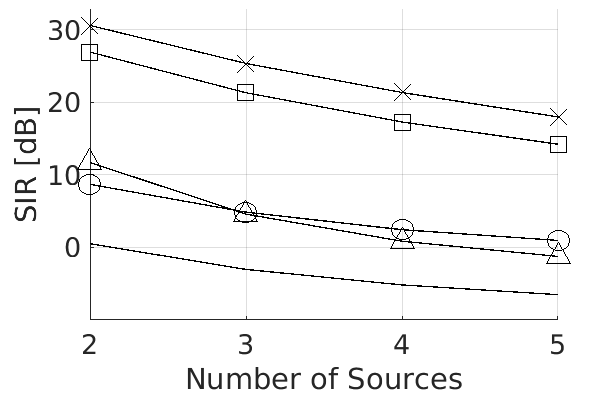}} 
{\includegraphics[width=0.31\textwidth]{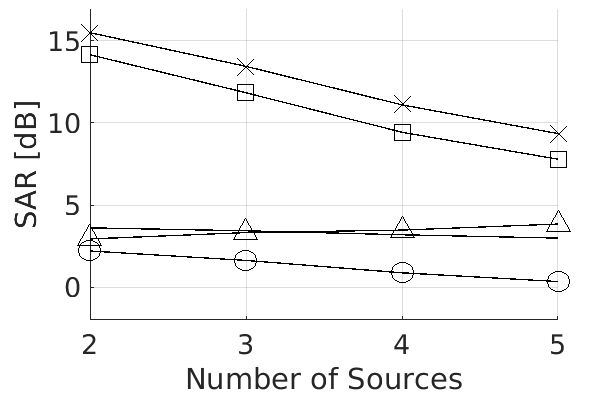}}  
\caption{Performance measures as a function of the number of sources. $T_{60}$ = 0.5 s.  NPM is -35 dB. } 
\label{fig:ns}
\vspace{-0.0cm}
\end{figure*}

\begin{figure*}[t]
\centering
{\includegraphics[width=0.31\textwidth]{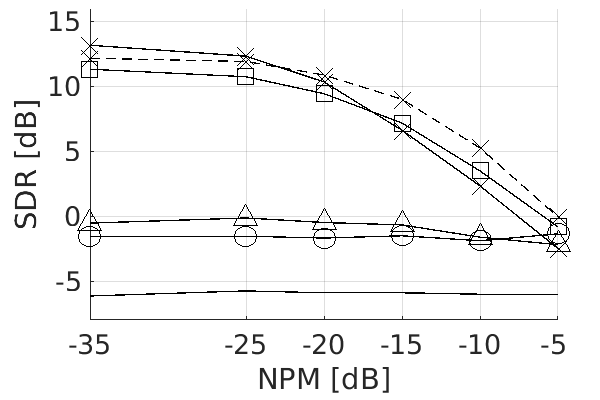}}
{\includegraphics[width=0.31\textwidth]{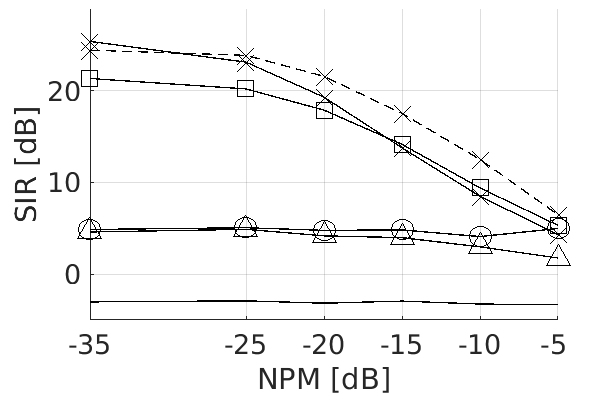}} 
{\includegraphics[width=0.31\textwidth]{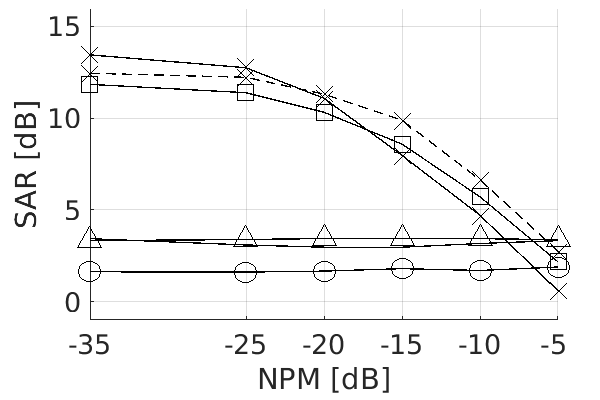}} 
\caption{Performance measures of the proposed methods as a function of NPM. $T_{60}$ = 0.5 s. The number of sources is 3. The solid curves with cross marker represent the proposed method with fixed CTFs, while the dashed curves with cross marker represent the proposed method with CTFs updated during EM iterations.} 
\label{fig:npm}
\vspace{-0.0cm}
\end{figure*}

Fig.~\ref{fig:t60} plots the performance measures obtained for the $4$ reverberation times. The number of sources is $3$. NPM is set to -35 dB, for which the filter pertrubation is light, and the CTFs used for CTF-Lasso and the proposed method are very accurate. Thence, for this experiment, the CTFs are fixed during EM iterations.  
For the anechoic case, all the four methods largely improve the SDR and SIR scores compared to the unprocessed signals, however slightly reduces the SAR score. This indicates that the three methods can efficiently separate the multiple sources, but introduces some artifacts. Especially, BM suffers from more artifacts than the other methods, since the hard assignment of the time-frequency bins to the dominant source largely distorts the less dominant sources.  
As $T_{60}$ increases, the performance measures of BM and FR-SCM dramatically decrease, since the length of RIRs is much larger than the STFT window and the narrowband approximation is not valid any more for the high reverberation cases. 
FR-SCM outperforms BM due to the use of the full-rank convariance matrix, which however is only suitable for the low reverberation cases. This can be testified by the fact that the most prominent advantage of FR-SCM over BM presents when $T_{60}$ is 0.22 s.  In contrast to BM and FR-SCM, CTF-Lasso and the proposed EM method achieve good performances. It is a bit surprising that the performance measures actually increases with the increase of reverberation time. The possible reason is that the long filters involve more information to differentiate and separate the multiple sources. Of course, to satisfy this assertion the filters should be accurate enough.  Compared to CTF-Lasso whose outputs are taken as the initial point for the EM algorithm, the EM   algorithm  improves the SDR by about 1.5~dB for every reverberation times, which indicates that the EM iterations are able to refine the quality of the source estimate.

\subsection{Results as a Function of Number of Sources}
Fig.~\ref{fig:ns} plots the results for various number of sources, for $T_{60}=0.5$~s. NPM is set to -35 dB, and the CTFs are fixed during EM iterations.  
As expected, the performance measures of all methods degrade when the number of sources increases. 
For BM, the WDO assumption for speech sources becomes less valid when more sources present.  For FR-SCM, CTF-Lasso and the proposed EM method, the mutual confusion of sources gets larger with increasing number of sources. The performance degradation rate of the four methods are similar to the one for the unprocessed signals. Overall, when the CTFs are properly initialized, CTF-Lasso and the proposed EM method achieve good source separtion performance even for the mixtures with five sources using only two microphones.

\subsection{Results as a Function of NPM}
To evaluate the proposed method under the conditions that the initialized CTFs suffer from large estimation error, we conducted the experiments with various NPM settings, and the results are shown in Fig.~\ref{fig:npm}. Since the initialization is not accurate, in this experiment, the CTFs are updated during EM interations to refine the CTFs. To demonstrate the efficiency of CTF updation, the results with the fixed CTFs are also given.   As NPM increases, the performance of CTF-Lasso and the proposed EM method largely degrade. When NPM is larger than -20 dB, the proposed method with fixed CTFs does not improve the performance of CTF-Lasso, which means the source estimation can not be refined based the inaccurate CTFs. In contrast, the proposed method with updating CTFs improves the performance of CTF-Lasso, due to the refining of CTFs in the M-step. The performance measures of CTF-Lasso and the proposed method become close to the ones of BM and FR-SCM, which indicates that the CTF-based methods are more sensitive to the filter perturbations than the narrowband assumption based methods.   

\subsection{Computational Complexity Analysis}

Table \ref{tab:rtf} shows the real-time factor of the four methods for the case with three sources and $T_{60}$ of 0.5 s. BM is the fastest, since it is an one-step method. The other three are all iterative methods, whose computational complexity is proportional to the number of iterations.  FR-SCM and the proposed method are similar in the sense that they both are based on EM iteration: estimating the source statistics using a Wiener-like filter in the E-step, and estimating the mixing filters in the M-step. The main difference between them is that the proposed method uses the CTF with a length of $Q$, while FR-SCM uses the mixing filter with a length of $1$. As a result, the proposed method has a much larger complexity than FR-SCM. CTF-Lasso also uses the CTF. However, unlike FR-SCM and the proposed method that several matrix inverse operations are performed (as shown in (\ref{eq:poss}) and (\ref{eq:Anew})), Lasso optimization only executes the first-order convolution operation. Thence, the real-time factor of CTF-Lasso is lower than the factor of FR-SCM and the proposed method. 

\begin{table}[t]
\centering
\caption{\small Real-time factor of the four methods.}
\label{tab:rtf}
\begin{tabular}{ c   c c  c    }   \hline
 BM   & FR-SCM  & CTF-Lasso   & Proposed  \\  
 0.01  &  81.7   & 54.2 & 630.0 \\ \hline
\end{tabular}
\end{table}     

\section{Conclusion}
\label{sec:con}

In this work, an EM algorithm has been proposed for speech separation. The subband convolutive model, i.e. CTF model, was adopt. To concisely derive the M-step and E-step, two convolution vector/matrix formulations were used. 
The CTF model based methods, i.e. CTF-Lasso and the proposed EM method, outperform the narrowband assumption based method, i.e. binary masking and FR-SCM, for the reverberant case. The proposed EM algorithm is capable to refine the CTFs and source estimates  starting with the output of CTF-Lasso, and thus improves the source separation performance. Only the semi-blind experiments were conducted in this work due to the difficulty of EM initialization. In the future, a blind CTF identification method could be developed to enable the blind initialization of EM. To this aim, the  CTF identification methods proposed in [23] and [30] could be combined, which are in the contexts of single-source dereverberation and multisource localization, respectively.

\section*{References}\label{sec:ref}

\begin{enumerate}
\item[{[1]}] Yilmaz, O., Rickard, S.: `Blind separation of speech mixtures via time-frequency masking', IEEE Transactions on Signal Processing, 2014,  52, (7), pp. 1830--1847 \vspace*{6pt}

\item[{[2]}]  Mandel, M. I., Weiss, R. J.,  Ellis, D. P.: `Model-based expectation-maximization source separation and localization', IEEE Transactions on Audio, Speech, and Language Processing, 2010, 18, (2), pp. 382--394 \vspace*{6pt}

\item[{[3]}] Winter, S., Kellermann, W., Sawada, H., Makino, S.: `MAP-based underdetermined blind source separation of convolutive mixtures by hierarchical clustering and $\ell_1$-norm minimization', EURASIP Journal on Applied Signal Processing, 2007, (1), pp. 81--81 \vspace*{6pt}

\item[{[4]}] Ozerov, A., F{\'e}votte, C.: `Multichannel nonnegative matrix factorization in convolutive mixtures for audio source separation', IEEE Transactions on Audio, Speech, and Language Processing, 2010, 18, (3), pp. 550--563 \vspace*{6pt}

\item[{[5]}] Avargel, Y., Cohen, I.: `On multiplicative transfer function approximation in the short-time Fourier transform domain', IEEE Signal Processing Letters, 2007, 14, (5), pp. 337--340 \vspace*{6pt}

\item[{[6]}] Gannot, S., Burshtein, D.,  Weinstein, E.: `Signal enhancement using beamforming and nonstationarity with applications to speech', IEEE Transactions on Signal Processing, 2001, 49, (8), pp. 1614--1626 \vspace*{6pt}

\item[{[7]}] Li, X., Girin, L., Horaud, R., Gannot, S.: `Estimation of relative transfer function in the presence of stationary noise based on segmental power spectral density matrix subtraction'.  in IEEE International Conference on Acoustics, Speech and Signal Processing, 2015, pp. 320--324 \vspace*{6pt}

\item[{[8]}]
Van Trees, H. L.: `Detection, estimation, and modulation theory' (John Wiley \&
Sons, 2004) \vspace*{6pt}

\item[{[9]}] Gannot, S., Vincent, E., Markovich-Golan, S., Ozerov, A.: `A consolidated perspective on multimicrophone speech enhancement and source separation', IEEE/ACM Transactions on Audio, Speech, and Language Processing, 2017, 25, (4), pp. 692--730 \vspace*{6pt}

\item[{[10]}] Duong, N., Vincent, E., Gribonval, R.: `Under-determined reverberant audio source separation using a full-rank spatial covariance model', IEEE Transactions on Audio, Speech, and Language Processing, 2010, 18, (7), pp. 1830--1840 \vspace*{6pt}

\item[{[11]}] Kowalski, M., Vincent, E., Gribonval, R.: `Beyond the narrowband approximation: Wideband convex methods for under-determined reverberant audio source separation', IEEE Transactions on Audio, Speech, and Language Processing, 2010, vol. 18, no. 7, pp. 1818--1829 \vspace*{6pt}

\item[{[12]}] Arberet, S., Vandergheynst, P., Carrillo, J.-P., Thiran, R. E.,  Wiaux, Y.: `Sparse reverberant audio source separation via reweighted analysis', IEEE Transactions on Audio, Speech, and Language Processing, 2013, 21, (7), pp. 1391--1402 \vspace*{6pt}

\item[{[13]}] Leglaive, S., Badeau, R.,  Richard, G.: `Multichannel audio source separation: variational inference of time-frequency sources from time-domain observations'. in IEEE International Conference on Acoustics, Speech and Signal Processing (ICASSP), 2017, pp. 26--30 \vspace*{6pt}

\item[{[14]}] Leglaive, S., Badeau, R.,  Richard, G.: `Separating time-frequency sources from time-domain convolutive mixtures using non-negative matrix factorization'. in IEEE Workshop on Applications of Signal Processing to Audio and Acoustics (WASPAA), 2017, pp. 264--268 \vspace*{6pt}

\item[{[15]}] Avargel, Y., Cohen, I.: `System identification in the short-time Fourier transform domain with crossband filtering', IEEE Transactions on Audio, Speech, and Language Processing, 2007, 15, (4), pp. 1305--319 \vspace*{6pt}

\item[{[16]}] Talmon, R., Cohen, I.,  Gannot, S.: `Relative transfer function identification using convolutive transfer function approximation', IEEE Transactions on Audio, Speech, and Language Processing, 2009, 17, (4), pp. 546--555  \vspace*{6pt}

\item[{[17]}] Li, X., Girin, L., Horaud, R.: `Audio source separation based on convolutive transfer function and frequency-domain lasso optimization'. in IEEE International Conference on Acoustics, Speech, and Signal Processing (ICASSP), 2017, pp. 541--545 \vspace*{6pt}

\item[{[18]}] Li, X., Girin, L., Gannot, S., Horaud, R.: `Multichannel speech separation and enhancement using the convolutive transfer function'. IEEE/ACM Transactions on Audio, Speech, and Language Processing, 2018 \vspace*{6pt}

\item[{[19]}] Talmon, R., I. Cohen, I., Gannot, S.: `Convolutive transfer function generalized sidelobe canceler', IEEE transactions on audio, speech, and language processing, 2009, 17, (7), pp. 1420--1434 \vspace*{6pt}

\item[{[20]}] Schwartz, B., Gannot, S., Habets, E. A.: `Online speech dereverberation using kalman filter and EM algorithm', IEEE/ACM Transactions on Audio, Speech and Language Processing, 2015, 23, (2), pp. 394--406 \vspace*{6pt}

\item[{[21]}] Badeau, R., Plumbley, M. D.: `Multichannel high-resolution NMF for modeling convolutive mixtures of non-stationary signals in the time-frequency domain', IEEE/ACM Transactions on Audio, Speech, and Language Processing, 2014,  22, (11), pp. 1670--1680 \vspace*{6pt}

\item[{[22]}] Higuchi, T., and Kameoka, H.: `Joint audio source separation and dereverberation based on multichannel factorial hidden Markov model', in IEEE International Workshop on Machine Learning for Signal Processing (MLSP), 2014, pp. 1--6 \vspace*{6pt}

\item[{[23]}] Li, X., Gannot, S., Girin, L.,  Horaud, R.: `Multichannel identification and nonnegative equalization for dereverberation and noise reduction based on convolutive transfer function', IEEE/ACM Transactions on Audio, Speech, and Language Processing, 2018, 26, (10), pp. 1755--1768 \vspace*{6pt}

\item[{[24]}] Li, X., Girin, L.,  Horaud, R.: `An EM algorithm for audio source separation based on the convolutive transfer function', in IEEE Workshop on Applications of Signal Processing to Audio and Acoustics (WASPAA), 2017, pp. 56--60 \vspace*{6pt}

\item[{[25]}] Gardner, W. G., Martin,  K. D.: `HRTF measurements of a KEMAR dummy-head microphone', The Journal of the Acoustical Society of America, 1995, 97, (6), pp. 3907--3908 \vspace*{6pt}

\item[{[26]}] Campbell, D.: `The roomsim user guide (v3.3)', https://pims. grc.nasa.gov/plots/user/acoustics/roomsim/Roomsim\%20User\%20 \\
Guide\%20v3p3.htm, 2004. \vspace*{6pt}

\item[{[27]}] Garofolo, J. S., Lamel, L. F., Fisher, W. M., Fiscus, J. G.,  Pallett, D. S., Dahlgren, N. L.: `Getting started with the DARPA TIMIT CD-ROM: An acoustic phonetic continuous speech database', National Institute of Standards and Technology  (NIST), Gaithersburgh, MD, 1988, 107 \vspace*{6pt}

\item[{[28]}] Morgan, D. R., Benesty, J., Sondhi, M. M.: `On the evaluation of estimated impulse responses', IEEE Signal processing letters, 1998, 5, (7), pp. 174--176 \vspace*{6pt}

\item[{[29]}] Vincent, E., Gribonval, R., F{\'e}votte, C.: `Performance measurement in blind  audio source separation', IEEE transactions on audio, speech, and language processing, 2006, 14, (4), pp. 1462--1469 \vspace*{6pt}

\item[{[30]}] Li, X.,  Girin, L., Horaud, R., Gannot, S.: `Multiple-speaker localization based on direct-path features and likelihood maximization with spatial sparsity regularization', IEEE/ACM Transactions on Audio, Speech, and Language Processing, 2017, 25, (10), pp. 1997--2012 \vspace*{6pt}

\end{enumerate}

\vfill\pagebreak

\end{document}